\documentclass[prl,aps,showpacs,twocolumn,floatfix]{revtex4-1}
\usepackage{amsmath, amsfonts, amssymb, bm, graphicx, color}

\newcommand{\bS}{{\bm S}}

\newcommand{\bsig}{{\bm\sigma}}

\newcommand{\e}{{\rm e}}
\newcommand{\ii}{{\rm i}}
\newcommand{\dd}{d^\dag}
\newcommand{\ed}{e^\dag}
\newcommand{\cd}{c^\dag}
\newcommand{\fd}{f^\dag}
\newcommand{\gd}{g^\dag}
\newcommand{\ua}{{\uparrow}}
\newcommand{\da}{{\downarrow}}
\newcommand{\vac}{\vert {\rm vac} \rangle}

\begin{document}

\title{Heavy-fermion valence-bond liquids in ultracold atoms: Cooperation of
 Kondo effect and geometric frustration}
\author{L. Isaev}
\author{A. M. Rey}
\affiliation{JILA, NIST \& Department of Physics, University of Colorado,
	440 UCB, Boulder, CO 80309, USA}

\begin{abstract}

 We analyze a microscopic mechanism behind coexistence of a heavy Fermi liquid
 and geometric frustration in Kondo lattices.
 We consider a geometrically frustrated periodic Anderson model and demonstrate
 how orbital fluctuations lead to a Kondo-screened phase in the limit of
 extreme strong frustration when only local {\it singlet} states participate in
 the low-energy physics.
 We also propose a setup to realize and study this exotic state with
 $SU (3)$-symmetric alkaline-earth cold atoms.

\end{abstract}

\pacs{71.27.+a, 75.10.Kt, 67.85.-d, 37.10.Jk}
\maketitle

\paragraph*{Introduction.}

Geometric lattice frustration plays a crucial role in Mott insulators
\cite{diep-2004-1} where it usually suppresses long-range magnetism by
enhancing the number of competing magnetic ground states. At zero temperature,
this degeneracy may be relieved in favor of a quantum non-magnetic phase such
as a spin liquid or valence bond ordering \cite{lacroix-2011-1}. On the
contrary, lattice topology in most metals is less important due to long-range
magnetic interactions mediated by the itinerant electrons and small static
magnetic moments.

The situation is different in cases when magnetic and itinerant behaviors
originate from physically distinct degrees of freedom \cite{kee-2012-1}. For
example, in heavy-fermion (HF) metals \cite{hewson-1997-1,coleman-2007-1}
magnetic
moments arise from localized $4f$ or $5f$-electrons, while conduction electrons
typically reside in extended atomic $s$-orbitals. Low-temperature properties of
such systems are driven by several opposing quantum many-body effects: (i)
Kondo screening, i.e. formation of singlets between local moments and itinerant
electrons that gives rise to ``heavy'' quasiparticle states with delocalized
$f$-electrons; (ii) local-moment long-range magnetism; and (iii) non-magnetic
states due to lattice frustration that involve singlets only among local spins.
Geometrically frustrated $f$-electron compounds [or Kondo lattices (KLs)] such
as ${\rm Yb_2Pt_2Pb}$ \cite{kim-2013-1} received much attention in the recent
years \cite{coleman-2010-1,si-2013-1,pixley-2014-1,bernhard-2011-1}.

The magnetism, Kondo effect, and geometric frustration compete because they
involve same local electrons which can not simultaneously form
singlets with each other and the conduction band. This observation is at the
heart of the recently proposed generic phase diagram of HF materials
\cite{coleman-2010-1} that allows their classification
according to the amount of quantum fluctuations of local
magnetism \cite{custers-2012-1}. Naturally, this phase diagram precludes Kondo
screening in strongly-frustrated lattices.
The antagonism between Kondo effect and lattice frustration only occurs in
cases that involve pure spin degrees of freedom. In contrast, in systems with
multiple local orbitals, orbital fluctuations allow local {\it spin singlets}
to participate in the Kondo screening
\cite{kiselev-2006-1,isaev-2013-1} together with the usual
``spinful'' states. If the singlets were due to frustration, the local orbital
fluctuations might provide a pathway towards a strongly-frustrated
Kondo-screened state.

\begin{figure}[t]
 \begin{center}
  \includegraphics[width = \columnwidth]{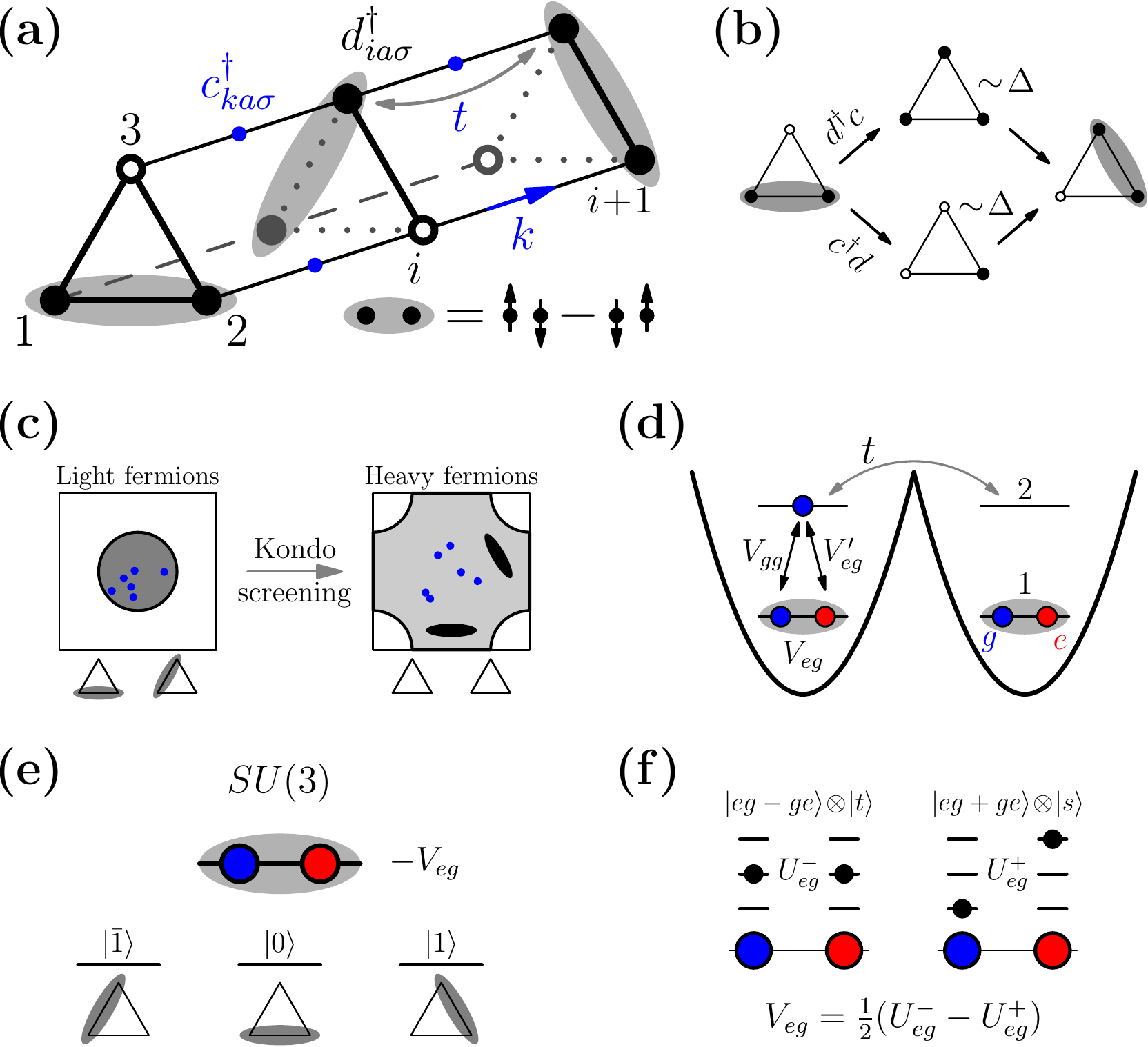}
 \end{center}
 \caption{
  Panel (a) The triangular tube lattice. Black and blue circles denote local
  ($d^\dag_{i a \sigma}$) and itinerant ($c^\dag_{k a \sigma}$) electrons. Grey
  ellipses are VB singlets \eqref{eq:vb-singlets} (empty circles indicate
  holes). Itinerant fermions propagate (by hopping between triangles with an
  amplitude $t$) in the leads with momentum $k$.
  (b) Valence fluctuations away from the two-electron singlet GS of a triangle,
  leading to VB flips.
  (c) Schematic plot of delocalization of the VB singlets and
  heavy-quasiparticle formation due to the Kondo screening. Shaded regions are
  Fermi surfaces in the Brillouin zone.
  (d) Magic optical lattice that implements the frustrated KL model
  \eqref{eq:klm-ttl} [band 1 (2) is localized (itinerant)]. Red and blue
  circles are AEAs in ${}^3 P_0$ ($e$) and ${}^1 S_0$ ($g$) clock states.
  Grey ellipses show local $e$-$g$ entangled states (with energy $-V_{e g}$).
  $V^\prime_{e g}$ and $V_{g g}$ are the inter-band exchange interactions
  ($V_{e g} \gg V^\prime_{e g}, \vert V_{g g} \vert$).
  (e) Mapping from VB singlets on a triangle to lowest-energy $e$-$g$ pair
  states for $SU(3)$ AEAs.
  (f) Two different scattering lengths for spin symmetric $\vert t \rangle$
  and antisymmetric $\vert s \rangle$ channels. The three levels indicate
  nuclear spin states for each atom (black circle marks a populated state).
 }
 \label{fig:fig1}
\end{figure}

In the present Letter we argue that such phase with coexisting Kondo and
frustration-driven local-spin singlets can indeed be realized.
To demonstrate this, we consider a toy system -- a periodic Anderson model on a
triangular tube lattice (TTL) of Fig. \ref{fig:fig1}(a) with frustrated
triangular plaquettes (due to large antiferromagnetic exchange interaction
between localized electrons) in the Kondo regime when valence fluctuations are
suppressed and each plaquette has a spin-singlet ground state (GS) with exactly
two fermions.
Because of different possible arrangements of local {\it valence bond} (VB)
singlets \cite{anderson-1988-1}, this GS is {\it triply degenerate}.
Although local spins are quenched in the singlet states, orbital fluctuations
[Fig. \ref{fig:fig1}(b)] allow mixing of the VB configurations by the Anderson
hybridization with the conduction band, and give rise to a robust
Kondo-screened GS with heavy quasiparticles and delocalized VB singlets [Fig.
\ref{fig:fig1}(c)].

This KL can be implemented using fermionic alkaline-earth atoms (AEAs), i.e.
atoms with two outer electrons, in an optical lattice [see Fig.
\ref{fig:fig1}(d)].
AEAs prepared in the two lowest clock states (${}^1 S_0$ and ${}^3 P_0$) with
total angular momentum $J = 0$ show a strict decoupling of electronic orbital
and nuclear-spin degrees of freedom, and obey an accurate $SU (N \leqslant 2 I
+ 1)$ ($I$ is the nuclear spin) symmetry in the two-body collisions
\cite{gorshkov-2010-1} which has been recently verified with
${}^{87}{\rm Sr}$ \cite{zhang-2014-1} and ${}^{173}{\rm Yb}$
\cite{cappellini-2014-1,scazza-2014-1}.
Our key observation is that the local VB singlets can be encoded with entangled
states of two AEAs [Fig. \ref{fig:fig1}(e)] prepared in different clock
configurations and three nuclear spin levels.
The degeneracy of these states is {\it guaranteed by the $SU (N = 3)$
symmetry}.
The entangled atomic pairs are loaded in the lowest, strongly localized, band
of a {\it magic} optical lattice whose trapping potential does not affect clock
transitions \cite{katori-2003-1}, and {\it implement the locally frustrated
plaquettes} (the optical lattice itself does not need to be geometrically
frustrated).
The conduction electrons are simulated by placing AEAs in a higher, itinerant
band.

At low energies, both of the above systems are described by a KL model with a
peculiar $SU (3)$ structure.
In the metallic regime, its GS is a Fermi [in one dimension (1D), Luttinger]
liquid consisting of delocalized VB singlets (AEA pairs) screened by itinerant
fermions, that can be viewed as a short-range resonant VB spin liquid
\cite{balents-2010-1} {\it stabilized by the Kondo effect}.

\paragraph*{Toy model: $SU (3)$ KL on a TTL.}

Let us consider a periodic Anderson model on the lattice of Fig.
\ref{fig:fig1}(a):
\begin{align}
 H^{TTL} = & \sum_{k a \sigma} \varepsilon_{k \sigma} \cd_{k a \sigma}
 c_{k a \sigma} + v \sum_{i a \sigma} (\cd_{i a \sigma} d_{i a \sigma} +
 {\rm h .c}) + \label{eq:pam-ttl} \\
 + \sum_i & \biggl[ J_H \bS_d^2(x_i) - \epsilon_d N_d(x_i) + U \sum_{a \neq b}
 n^d_{i a} n^d_{i b} \biggr] + H_{\rm mix},
 \nonumber
\end{align}
which describes a system of conduction electrons $c_{k a \sigma}$ with momentum
$k$ in the $a$th lead ($a = 1 \ldots 3$), spin $\sigma = \lbrace \ua, \da
\rbrace$, hybridized (via an amplitude $v$) with
local electrons $d_{i a \sigma}$ at each vertex $a$ of a triangle at position
$x_i = i$. $N_d =
\sum_a n^d_{i a} = \sum_{a \sigma} \dd_{i a \sigma} d_{i a \sigma}$ and
$\bS_d = \frac{1}{2} \sum_{a \alpha \beta} \dd_{i a \alpha}
\bsig_{\alpha \beta} d_{i a \beta}$
($\bsig$ are Pauli matrices) define electron number and total spin of a
triangle. The
dispersion $\varepsilon_{k \sigma} = \epsilon_k - h \sigma$ includes a small
(compared to other magnetic interactions) Zeeman splitting $h$ whose role we
explain later.
The term $H_{\rm mix}$ describes mixing of fermions in different leads $a$ and
for now will be ignored.

There are several energy scales associated with each triangle: local binding
energy $\epsilon_d > 0$, the nearest-neighbor Coulomb repulsion $U$, ``Hund''
energy $J_H \geqslant 0$ that forces the lowest total spin $S_d$, and an
infinitely large on-site Coulomb repulsion preventing double occupancy of any
vertex $a$. We focus on a two-electron $S_d = 0$ subspace which contains a
three-fold degenerate GS when $U - \frac{3}{4} J_H < \epsilon_d < 2 U +
\frac{3}{4} J_H$:
\begin{equation}
 \vert a \rangle_i = {\textstyle \frac{1}{\sqrt{2}}} \sideset{}{_{b^\prime b}}
 \sum s^a_{b^\prime b} \dd_{i b^\prime \ua} \dd_{i b \da} \vac,
 \label{eq:vb-singlets}
\end{equation}
where $s^a_{b^\prime b} = s^a_{b b^\prime} = 1$ when $a$, $b$ and $b^\prime$
are different, and $0$ otherwise; $\vac$ is the vacuum ($N_d = 0$) state. These
states are labeled by the number of an unoccupied vertex.

We will fix $\epsilon_d = \frac{3}{2} U$ and consider the strong-coupling
regime $v \ll \epsilon_d,\, U,\, J_H$ when $N_d$-fluctuations on each
triangle are virtual and can be taken into account via a generalized
Schrieffer-Wolff transformation ${\cal S}$
\cite{[][{; see also the Supplementary material.}] muhlschlegel-1968-1} that
includes processes shown in Fig. \ref{fig:fig1}(b). A straightforward
calculation yields the second-order KL Hamiltonian
\begin{equation}
 H^{TTL}_{\rm ef} = \sum_{k a \sigma} \varepsilon_{k \sigma} \cd_{k a \sigma}
 c_{k a \sigma} - \sum_{i \sigma a b} V_{a b} \fd_{i a} f_{i b}
 \cd_{i a \sigma} c_{i b \sigma}
 \label{eq:klm-ttl}
\end{equation}
that describes scattering of conduction electrons by the local VB singlets and
is defined on a {\it non-frustrated} lattice whose sites correspond to
triangles in Fig. \ref{fig:fig1}(a).
The coupling constants are $V_{a b} = V_\perp (1 - \delta_{a b}) + V_\|
\delta_{a b}$ with $V_\perp = -\frac{v^2}{2 \Delta}$, $V_\| =
\frac{3 v^2}{2 \Delta}$, $\delta_{a b}$ -- the Kronecker delta, and the valence
fluctuation gap $\Delta = \frac{1}{2} U + \frac{3}{4} J_H$.
The states \eqref{eq:vb-singlets} are described with a pseudo-fermion
representation \cite{coleman-2007-1}:
\begin{equation}
 \vert a \rangle_i \to \fd_{i a} \vac
 \label{eq:pseudo-fermions}
\end{equation} 
with a Hilbert space constraint $\sum_a \fd_{i a} f_{i a} = 1$. Because only
$S_d = 0$ triangle states are involved in the low-energy physics, interactions
in $H^{TTL}_{\rm ef}$ preserve electron spin $\sigma$ and only change the
orbital (lead) degree of freedom $a$.

As a result,
Eq. \eqref{eq:klm-ttl} describes a two-channel KL model (spin is the channel
index) \cite{cox-1999-1}. It is known that the two-channel fixed point is
usually unstable w.r.t. channel asymmetry \cite{cox-1999-1} controlled by the
Zeeman splitting $h$. Since even for small $h \ll J_H$ the leads may be
considered spin-polarized, below we omit the spin index $\sigma$ and replace
$c_{i a \sigma} \to c_{i a \ua} \equiv c_{i a}$ and $\varepsilon_{k \sigma} =
\varepsilon_{k \ua} \approx \epsilon_k$.

The Hamiltonian \eqref{eq:klm-ttl} contains matrix elements connecting all
three possible local VB states and conduction electron ``flavors'' $a$, and is
an anisotropic ($XXZ$-like) $SU(3)$ KL model written in terms of generators
$T_a^b(x_i) = \fd_{i a} f_{i b}$ and ${\tilde \tau}_a^b(x_i) = \cd_{i b}
c_{i a}$ for local and itinerant degrees of freedom \cite{[{
 ${\tilde \tau}_a^b$ are conjugate generators (note the order of indices $a$
 and $b$). For more details see: 
}] auerbach-1994-1}.
The local $SU (3)$ ``spin'' operators $T_a^b (x_i)$ describe orbital
fluctuations in Eq. \eqref{eq:pam-ttl} that flip the VB singlets
\eqref{eq:vb-singlets}.
$H^{TTL}_{\rm ef}$ in Eq. \eqref{eq:klm-ttl} is invariant under $U (1)$
transformations $f_{i a} \to \e^{\ii \phi_a} f_{i a}$ and $c_{i a} \to
\e^{-\ii \phi_a} c_{i a}$ that preserve the $V_\perp$ term.
There is also a discrete lattice symmetry $C_{3 v} = \lbrace C_3, \, \sigma_v
\rbrace$ \cite{bir-1974-1} that contains $\frac{2 \pi}{3}$ ($C_3$) rotations
around the TTL axis and three symmetry planes $\sigma_v$ of the triangles.

\begin{figure}[b]
 \begin{center}
  \includegraphics[width = 0.9 \columnwidth]{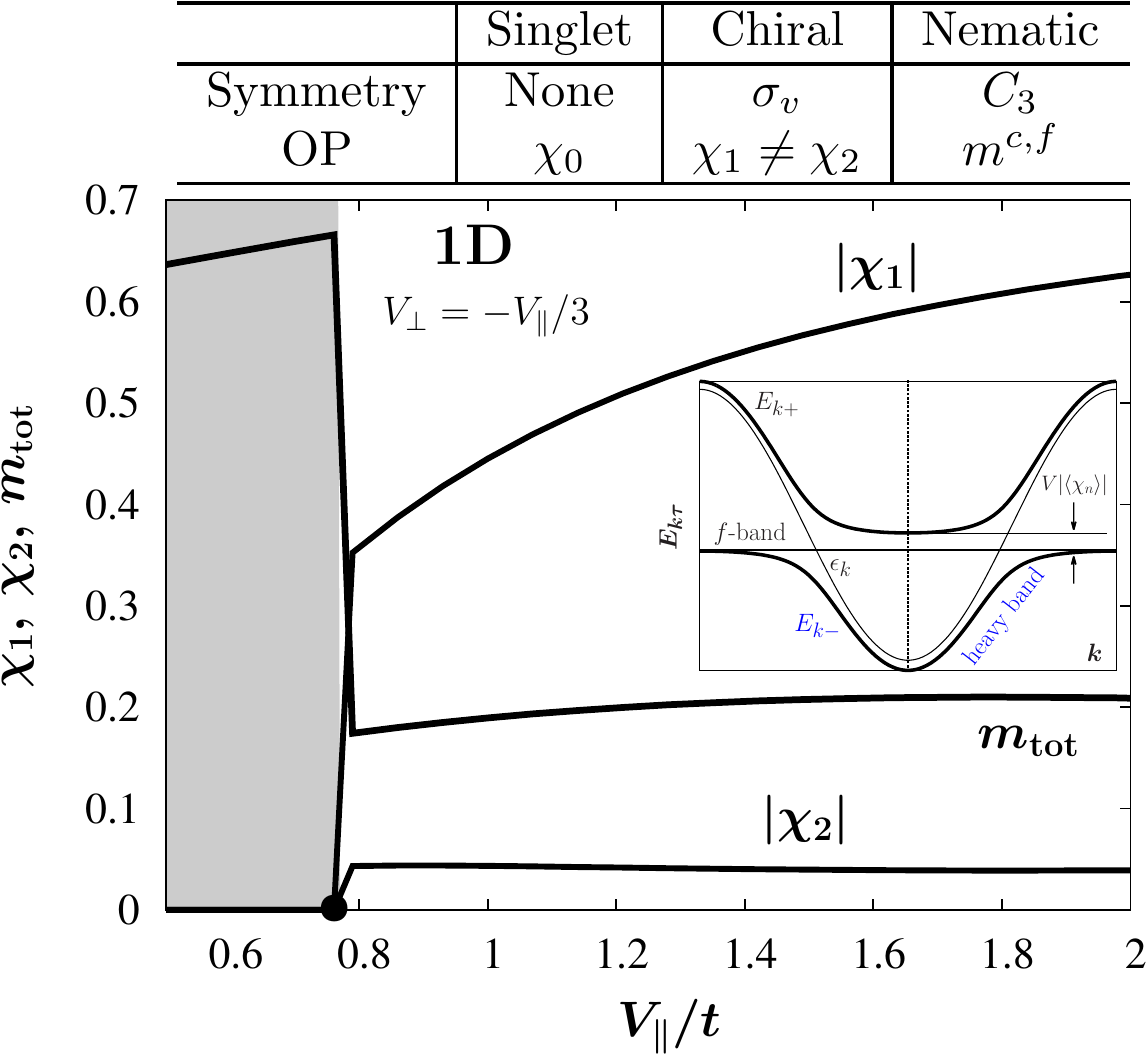}
 \end{center}
 \caption{
  $T = 0$ phase diagram of the KL model \eqref{eq:klm-ttl} with $N = 4900$
  sites and electron density $n^c = 0.8$.
  $m_{\rm tot} = [(m^c_3 + m^f_3)^2 + (m^c_8 + m^f_8)^2]^{1/2}$ plays a role of
  the total magnetization.
  The black circle at $V_\| /t \sim 0.76$ marks the first order transition
  between Kondo-screened and normal phases.
  Inset: The HF band-structure.
  The band splitting at the Fermi level is $\sim V \chi_{1, 2}$.
  The table shows symmetries broken by different OPs and corresponding VB
  liquid phases.
 }
 \label{fig:fig2}
\end{figure}

\paragraph*{Kondo effect-assisted VB phases.}

To demonstrate that the model \eqref{eq:klm-ttl} has a Kondo screened GS, we
use a generalized hybridization mean-field (HMF) approach
\cite{kusminskiy-2008-1} that treats the $f$-fermion Hilbert space
constraint on average, $\frac{1}{N} \sum_{i a} \langle \fd_{i a} f_{i a}
\rangle = 1$ ($N$ is the system size), and self-consistently compute the
hybridization and $SU(3)$ ``magnetization'' order parameters (OPs) \footnote{
 See the Supplementary material
}.
We assume that all OPs are site-independent.
There are three hybridization amplitudes:
$\chi_0 = \frac{1}{\sqrt{3}} \sum_a \langle f_{i a} c_{i a} \rangle$,
$\chi_{1, 2} = \frac{1}{\sqrt{3}} \langle f_{i 1} c_{i 1} + \omega^{\mp 1}
f_{i 2} c_{i 2} + \omega^{\pm 1} f_{i 3} c_{i 3} \rangle$ with $\omega =
\e^{2 \pi \ii / 3}$, and eight magnetizations $m^c_l$ [$m^f_l$] for $c$- [$f$-]
fermions defined via $\langle \cd_{i a} c_{i b} \rangle = \sum_l
\lambda^l_{a b} m^c_l + \frac{n^c}{3} \delta_{a b}$ [$\langle \fd_{i a} f_{i b}
\rangle = \sum_l \lambda^l_{a b} m^f_l + \frac{1}{3} \delta_{a b}$] where
$\lambda^l$ are the Gell-Mann matrices, $l = 1 \ldots 8$ and $n^c$ is the
conduction band filling.
Unlike the real $SU (2)$ magnetization, $m^{c,f}$ do not break time-reversal
invariance but rather the above $U (1)$ and $C_{3 v}$ symmetries.
The OPs $\chi_1$ and $\chi_2$ are connected (up to a phase) by the planes
$\sigma_v$ from $C_{3 v}$.
Finite $m^{c, f}_{3, 8}$ completely break $C_{3 v}$ leading to nematic states;
$m^{c, f}_l$ with $l \neq 3,\, 8$ also break the above $U (1)$ symmetry. 

Kondo-screened states correspond to nonzero values of either hybridization OP
$\chi_{0, 1, 2}$.
In analogy to spin systems \cite{okunishi-2012-1,seki-2015-1}, we call phases
with $\chi_1 \neq \chi_2$ chiral \footnote{
 Our notion of chirality is similar to the vector chirality in spin systems
 ${\bm \kappa} \sim [\bS_1 \times \bS_2$] \cite{okunishi-2012-1}. Indeed, for a
 given triangle, one can introduce an analogous quantity $\kappa_a = \ii
 \sum_{b^\prime b} \varepsilon_{a b^\prime b} \fd_{b^\prime} \cd_{b^\prime} c_b
 f_b$
}.
This discussion is summarized in the table in Fig. \ref{fig:fig2}.

The phase diagram of the Hamiltonian \eqref{eq:klm-ttl} is shown in Fig.
\ref{fig:fig2} for $\epsilon_k = -2 t \cos k$ ($t$ is the nearest-neighbor
hopping). There is a first order transition between a normal state with
$\chi_{0, 1, 2} = 0$, and a Kondo screened phase with $\chi_1 \neq \chi_2 \neq
0$ (but $\chi_0 = 0$) and non-zero $m^{c,f}_{3,8}$. This chiral nematic phase
has delocalized VB singlets. The OPs
$m^{c,f}$ survive only at low temperature $T \leqslant T_c \sim 5 \times
10^{-2} t$; for $T > T_c$ the only finite OP is $\chi_1$ and the GS realizes a
chiral metallic VB spin liquid.

This phase is quite different from the Kondo-stabilized spin liquid of Ref.
\onlinecite{coleman-1989-1} where the resonating VBs of local spins are formed
due to their coupling to conduction band and are unstable away from the Kondo
regime.
In our case the VB singlets are due to geometric frustration, and the Kondo
screening only injects them into the Fermi sea.

\paragraph*{Stability of the Kondo-assisted VB liquid.}

\begin{figure}[t]
 \begin{center}
  \includegraphics[width = 0.9 \columnwidth]{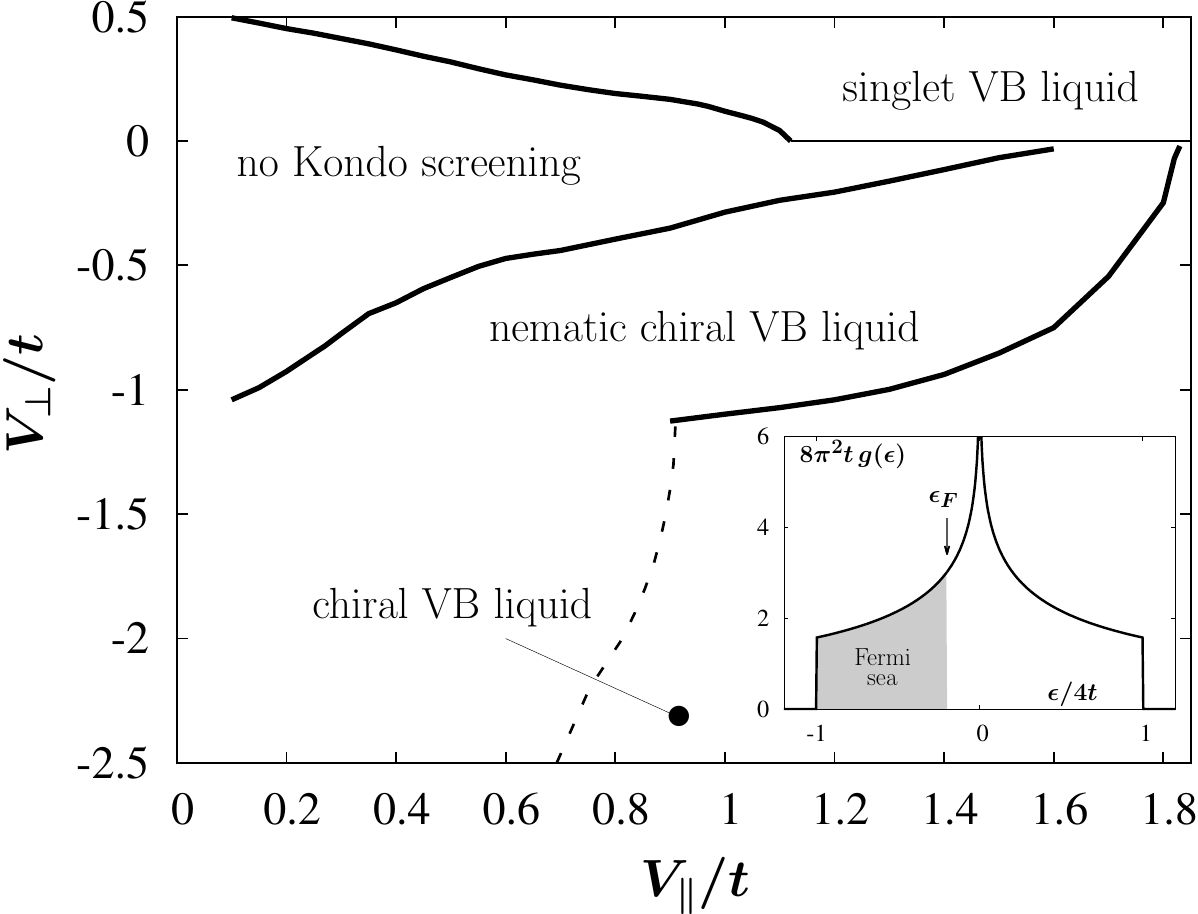}
 \end{center}
 \caption{
  Generic $T = 0$ phase diagram of Eq. \eqref{eq:klm-ttl} [or
  \eqref{eq:klm-aea} with $U_{g g} = 0$] with $N = 3600$ sites and $n^c = 0.8$.
  All phase transitions are first order.
  Inside the nematic phase there is a ``metamagnetic'' transition between
  states with $m^{c, f}_1 \neq 0$, $m^{c, f}_{3, 8} = 0$ (at smaller $V_\|$)
  and $m^{c, f}_1 = 0$, $m^{c, f}_{3, 8} \neq 0$ (for larger $V_\|$) which are
  separated by a continuation of the dashed line.
  Inset: Noninteracting itinerant DOS $g (\epsilon) = \frac{1}{8 \pi^2 t} K
  \bigl( \sqrt{1 - (\epsilon / 4t)^2} \bigr)$ [$K(x)$ is elliptic integral of
  the 1st kind] used to compute the phase diagram.
 }
 \label{fig:fig3}
\end{figure}

The Kondo phase in Fig. \ref{fig:fig2} is quite robust against changes in the
noninteracting itinerant density of states (DOS).
To show this, we consider a model DOS that corresponds to a square lattice with
nearest-neighbor hopping $\epsilon_k = -2 t (\cos k_x + \cos k_y)$ (as opposed
to the 1D tight-binding dispersion used before), see inset in Fig.
\ref{fig:fig3}.
The phase diagram obtained by applying the HMF approach to the KL
\eqref{eq:klm-ttl} is presented in Fig. \ref{fig:fig3}.
Unlike the 1D case in Fig. \ref{fig:fig2}, the chiral VB liquid with $\chi_1
\neq 0$, $\chi_{0, 2} = 0$ and $m^c = m^f = 0$ exists even at $T = 0$ for
$V_\perp \leqslant 0$ and large $V_\|$.
Only mirror symmetry from $C_{3 v}$ is broken by this state.
With decreasing $\vert V_\perp \vert$ and $V_\|$ the system undergoes a
transition to a nematic metallic state with $m^{c,f} \neq 0$ and completely
broken $C_{3 v}$.
The situation is different for $V_\perp \geqslant 0$.
Here the only non-zero OP is $\chi_0$ and the VB liquid GS does not break any
discrete symmetry.
All these Kondo-screened states become unstable at small $\vert V_\perp \vert$
and $V_\|$.

The phase transitions in Fig. \ref{fig:fig2} and \ref{fig:fig3} are first order
which may be an artifact of the HMF approximation.
In general at $T = 0$ the emergence of nonzero OPs $\chi_{0, 1, 2}$ is
associated with a {\it phase transition} (as opposed to a crossover) when
fluctuations beyond HMF are taken into account \cite{senthil-2004-1}.
Therefore salient features of our phase diagrams should remain unchanged.

Finally, we mention effects of a finite lead-mixing $H_{\rm mix}$ in Eq.
\eqref{eq:pam-ttl}.
Its simplest form (compatible with $C_{3 v}$ symmetry of the TTL) corresponds
to hopping of itinerant and local fermions around the triangle.
This correction results
\footnote{
 See the Supplementary material
}
in a Zeeman-like term, proportional to the intra-triangle hopping, which lifts
degeneracy of the local VB states \eqref{eq:vb-singlets} and can suppress the
Kondo phase in Fig. \ref{fig:fig2} if this splitting is sufficiently large
\cite{hewson-1997-1}.

\paragraph*{Implementation with ultracold AEAs.}
 
We propose an experimentally accessible implementation of the KL model
\eqref{eq:klm-ttl} with AEAs in an optical lattice, {\it that is free of the
mixing} described by $H_{\rm mix}$.
The key idea of our approach is
to use nuclear spin states of the atoms as ``synthetic'' frustrated plaquettes
[corresponding to triangles in Fig. \ref{fig:fig1}(a)] and construct an
appropriate low-energy model that takes into account these local states as well
as the itinerant degrees of freedom, and is unitarily related to the KL model
\eqref{eq:klm-ttl}.
The GS degeneracy of a synthetic plaquette is guaranteed by the $SU (N = 3)$
symmetry of the AEAs.

Consider a two-band optical lattice schematically shown in Fig.
\ref{fig:fig1}(d). The lowest-energy band is localized
and contains two AEAs per site in different clock states: one ${}^1 S_0$ (GS,
$g$) and one ${}^3 P_0$ (excited state, $e$). To minimize lossy $e$-$e$
collisions, the higher-energy itinerant band is populated only by $g$ atoms.
The Hamiltonian of the system is \cite{gorshkov-2010-1}:
\begin{align}
 & H^A = -t \sum_{\langle i j \rangle} (\cd_{i n} c_{j n} + {\rm h.c.}) +
 \sum_i \biggl[ \frac{U_{g g}}{2} n^c_i (n^c_i - 1) + \label{eq:sun-model} \\
 & \,\,\, + \bigl( V_{g g} \gd_{i n} g_{i m} + V^\prime_{e g} \ed_{i n}
 e_{i m} \bigr) \cd_{i m} c_{i n} + V_{e g} \ed_{i n} e_{i m} \gd_{i m}
 g_{i n} \biggr],
 \nonumber
\end{align}
where $g_{i n}$ ($e_{i n}$) denote $g$ ($e$) fermions in the localized band
at site $i$ and nuclear spin state $n = \bar{1},\, 0,\, 1$ [$\bar{n} = -n$,
i.e. $\bar{1} = -1$, $\bar{0} = 0$], and $c^\dag_{i n}$ create itinerant $g$
atoms.
There is an implicit summation over nuclear spin indices. The first term
describes nearest-neighbor hopping with an amplitude $t$. The second sum
corresponds to $e$-$g$ ($V_{e g}$ and $V^\prime_{e g}$) and $g$-$g$ ($V_{g g}$)
exchange couplings, as well as direct $g$-$g$ interaction $U_{g g} \geqslant 0$
[see Fig. \ref{fig:fig1}(d)-(f)]. $V_{e g}$ and $V^\prime_{e g}$ have the same
sign, and $V_{g g}$ is negative \footnote{
 Because the $s$-wave scattering length is $a_{g g} > 0$
 \cite{gorshkov-2010-1}, two-atom collisions favor antisymmetric spatial
 wavefunction and symmetric nuclear spin configurations.
}.

States of a localized $e$-$g$ pair are described by the term $H_{\rm loc}(x_i)
= V_{e g} \gd_{i n} g_{i m} \ed_{i m} e_{i n}$ whose spectrum consists of a
triply-degenerate GS subspace with energy $-V_{e g}$:
\begin{equation}
 \vert l \rangle_i = {\textstyle \frac{1}{\sqrt{2}}} \varepsilon_{l n m}
 \ed_{i n} \gd_{i m} \vac
 \label{eq:su3-singlets}
\end{equation}
($\varepsilon_{l n m}$ is the antisymmetric Levi-Civita tensor,
$\varepsilon_{1 0 \bar{1}} = 1$), and a sextet $\vert l \rangle^\prime_i = (1 /
\sqrt{2}) s^l_{n m} \ed_{i n} \gd_{i m} \vac$ [$s^a_{b c}$ was defined in
\eqref{eq:vb-singlets}] and $\vert l \rangle^{\prime \prime}_i = \ed_{i l}
\gd_{i l} \vac$ with energy $+V_{e g}$. We assume that $V_{e g}$ is large,
$V_{e g} \gg \vert V_{g g} \vert,\, V^\prime_{e g},\, t$ \footnote{
 $V_{e g}$ is at least twice larger than $V_{e g}^\prime$ and $|V_{g g}|$
 because
 of the Bloch-function overlap between lowest and excited bands. This overlap
 can be further decreased by placing itinerant $g$ atoms in higher bands
}, neglect mixing of the above sectors, and project the Hamiltonian
\eqref{eq:sun-model} onto the subspace \eqref{eq:su3-singlets}. Using the
relations
${}_i \langle l \vert \ed_{i n} e_{i m} \vert p \rangle_i = {}_i \langle l
\vert \gd_{i n} g_{i m} \vert p \rangle_i = \frac{1}{2} (\delta_{l p}
\delta_{n m} - \delta_{m l} \delta_{n p})$, and the pseudo-fermions
\eqref{eq:pseudo-fermions}, we obtain an effective model
\begin{equation}
 H^A_{\rm ef} = \sum_{k l} \epsilon_k \cd_{k l} c_{k l} - \sum_i \biggl[
 V \fd_{i l} f_{i p} \cd_{i l} c_{i p} - \frac{U_{g g}}{2} n^c_i (n^c_i - 1)
 \biggr]
 \label{eq:klm-aea}
\end{equation}
with $V = \frac{V^\prime_{e g} + V_{g g}}{2}$.
If the states \eqref{eq:su3-singlets} are identified with VB singlets
\eqref{eq:vb-singlets} on a triangle [Fig. \ref{fig:fig1}(e)] by assigning a
nuclear spin flavor $m$ to each vertex, $H^A_{\rm ef}$ in 1D is equivalent to
(spin-polarized) $H^{TTL}_{\rm ef}$ in Eq. \eqref{eq:klm-ttl} with $V_\perp =
V_\| = V$
\footnote{
 In the case of AEAs, $C_{3 v}$ operations should be applied to the
 nuclear spin states: $C_3$ rotations perform a cyclic permutation $1 0 \bar{1}
 \to \bar{1} 1 0$, the three mirror planes interchange any two states while
 preserving the third, e.g. $1 0 \bar{1} \to \bar{1} 0 1$
}
plus a Hubbard term, whose role as well as possible ways to introduce
anisotropic couplings in Eq. \eqref{eq:klm-aea} we discuss below.
To reach a Kondo screened GS one must have $V > 0$, i.e. $V^\prime_{e g} >
-V_{g g} > 0$ \footnote{
 See the renormalization group analysis in the Supplementary Material.
}.

\paragraph*{Discussion.}

Our theory highlights the fundamental role played by the orbital
degrees of freedom in stabilizing a {\it Kondo-screened phase} in the presence
of {\it extreme strong frustration} when only {\it singlet} local states
participate in the low-energy physics, by allowing the conduction electrons to
dynamically flip the VB singlets [see Fig. \ref{fig:fig1}(b)]. These
microscopic processes lead to delocalization of the local VBs and drive the
formation of the VB spin liquid with HF quasiparticles.
We illustrated this mechanism by studying a periodic Anderson model on a
frustrated triangular tube, and proposed a optical lattice setup to realize
this toy model with $SU (3)$-symmetric AEAs that employs
their nuclear-spin degrees of freedom to implement geometrically frustrated
plaquettes (e.g. triangles).

Compared to the electronic KL \eqref{eq:klm-ttl}, the low-energy model for AEAs
\eqref{eq:klm-aea} has several peculiarities.
First, there is the Hubbard term $U_{g g}$ which below half-filling enhances
phases with non-zero $SU (3)$ magnetization in Fig. \ref{fig:fig3}(b).
However, its magnitude is effectively damped by the density prefactor $\sim
(n^c)^2$.
We checked that even when $n^c = 0.8$, one needs $U_{g g} > V$ to suppress the
Kondo-screened state.
Hence this term is unimportant for the Kondo physics.
Second, the Hamiltonian $H^A_{\rm ef}$ has full $SU (3)$ symmetry (i.e.
$V_\perp = V_\|$) that originates from the symmetry of Eq. \eqref{eq:sun-model}
and prohibits experimental exploration of the phase diagram in Fig.
\ref{fig:fig3}.
This symmetry can be broken by a weak external magnetic field $B$ which to the
lowest order amounts to replacing $V_\perp \to V_\perp + \frac{V_{g g} -
V^\prime_{e g}}{V_{e g}} (\mu_e - \mu_g) B$ ($\mu_{e, g}$ are magnetic moments
for $e$ and $g$ atoms).
Also, one might use alternative implementations of the $SU(3)$ Kondo effect,
e.g. using orbital degrees of freedom \cite{nishida-2013-1}, instead of the
AEAs setup discussed here.

The HF phase in Figs. \ref{fig:fig2} and \ref{fig:fig3} can be detected in
cold-atom experiments using slow quantum dynamics or time-of-flight
measurements \cite{foss-feig-2010-1,foss-feig-2010-2,paredes-2005-1}.
The KL model in Eq. \eqref{eq:klm-aea} can be implemented beyond 1D, which
enables us to use AEAs as controlled [because of the $SU (N)$ symmetry] quantum
simulators for more complex frustrated Kondo lattices.
Although the currently available isotopes ${}^{87} {\rm Sr}$ and ${}^{173} {\rm
Yb}$ are believed to have negative exchange couplings $V$
\cite{zhang-2014-1,cappellini-2014-1,scazza-2014-1}, we expect that our
results summarized in Figs. \ref{fig:fig2} and \ref{fig:fig3} can be realized
with other AEAs.

\paragraph*{Acknowledgments.}

We are grateful to Gia-Wei Chern and Michael Hermele for illuminating
discussions. This work was supported by the NSF (PIF-1211914 and PFC-1125844),
AFOSR, AFOSR-MURI, NIST and ARO individual investigator awards.

\bibliographystyle{apsrev4-1}
\bibliography{references}
\end{document}


\title{Supplementary material for: Heavy-fermion valence-bond liquids in
ultracold atoms: cooperation of Kondo effect and geometric frustration}
\author{}
\affiliation{}
\begin{abstract}
\end{abstract}
\pacs{}
\maketitle

\section{$SU(3)$ KL model on a TTL}

In this section we outline a derivation of the KL Hamiltonian (3). We start
by considering an Anderson impurity model for an isolated triangle in Fig.
1(a), i.e. a single-rung case of Eq. (1). In the limit of weak valence
fluctuations, we use a Schrieffer-Wolff transformation to obtain a second-order
Kondo-like impurity Hamiltonian. Because the periodic Anderson model (1) is
additive w.r.t. triangular rungs, this Kondo impurity model can be
straightforwardly generalized for a periodic array of the ``impurities''
(triangles), i.e. a TTL.

The single-triangle Anderson impurity model describes three conduction bands
coupled to a triangular plaquette
\begin{align}
 H_{AIM} = & H_0 + H_{\rm int} + H_d; \label{eq:aim-3qd} \\
 H_0 = & \sum_{k a \sigma} \varepsilon_{k \sigma} \cd_{k a \sigma}
 c_{k a \sigma}; \nonumber \\
 H_{\rm int} = & \frac{v}{\sqrt{N}} \sum_{k a \sigma} (\cd_{k a \sigma}
 d_{a \sigma} + {\rm h .c}); \nonumber \\
 H_d = & J_H \bS_d^2 - \epsilon_d N_d + U \sum_{a \neq b} n^d_a n^d_b,
 \nonumber
\end{align}
where we used the same notations as in Eq. (1) but omitted the triangle
position $x_i = i$. $N$ is the dimensionless system size, e.g. the number of
unit cells.

\begin{figure}[b]
 \begin{center}
  \includegraphics[width = \columnwidth]{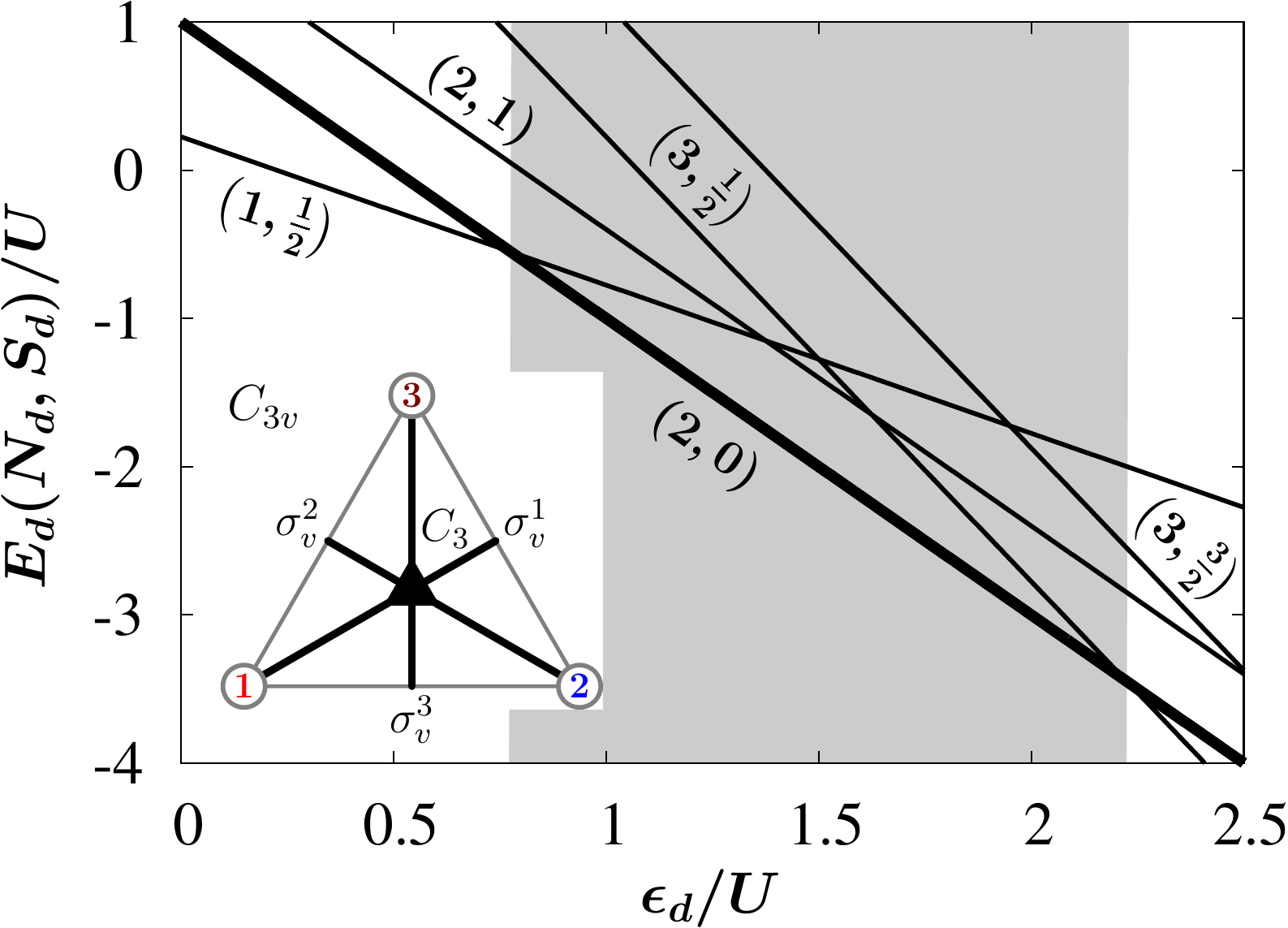}
 \end{center}
 \caption{
  Energy levels of an isolated triangle. $N_d$ and $S_d$ are the total particle
  number and spin. In the shaded region the GS has $N_d = 2$ electrons (thick
  line).
  Inset: The $C_{3 v}$ symmetry group of the triangle, which contains the
  three-fold axis $C_3$ and three mirror planes $\sigma_v$. The numbers $1
  \ldots 3$ indicate the vertices.
 }
 \label{fig:sfig1}
\end{figure}

The last line in Eq. \eqref{eq:aim-3qd} is the Hamiltonian of an isolated
triangle. Due to the strong on-site Coulomb repulsion we can treat
$d_{a \sigma}$ as constrained fermions (no double occupancy) with
anticommutation relations \cite{batista-2004-1} 
\begin{align}
 & \lbrace d_{a \alpha}, d_{b \beta} \rbrace = \lbrace \dd_{a \alpha}
 \dd_{b \beta} \rbrace = 0; \nonumber \\
 & \lbrace d_{a \alpha}, \dd_{b \beta} \rbrace = \delta_{a b} \bigl[
 \dd_{a \beta} d_{a \alpha} + \delta_{\alpha \beta} (1 - n^d_a) \bigr],
 \nonumber
\end{align}
and consider only three states per vertex $a$. There are hence $4^3 = 64$
electronic states. The corresponding energies $E_d(N_d, S_d)$,
\begin{align}
 E_d(1, 1/2) = & -\epsilon_d + \frac{3}{4} J_H, \nonumber \\
 E_d(2, 0) = & - 2 \epsilon_d + U, \nonumber \\
 E_d(2, 1) = & - 2 \epsilon_d + U + 2 J_H, \nonumber \\
 E_d(3, 1/2) = & - 3 \epsilon_d + 3 U + \frac{3}{4} J_H, \nonumber \\
 E_d(3, 3/2) = & - 3 \epsilon_d + 3 U + \frac{15}{4} J_H, \nonumber
\end{align}
are plotted in Fig. \ref{fig:sfig1}. For us the most interest has the
two-electron subspace with a singlet GS. This state is three-fold degenerate
and corresponds to a global energy minimum for $U - \frac{3}{4} J_H <
\epsilon_d < 2 U + \frac{3}{4} J_H$. The three singlet ground states are given
in Eq. (2).

In the strong-coupling regime $v \ll \epsilon_d,\, U$ we can use a
Schrieffer-Wolff transformation $H = \e^{\cal S} H_{AIM} \e^{-{\cal S}}$
\cite{hewson-1997-1} which is constructed to eliminate $H_{\rm int}$ in each
order in $v$. To the second order we have
\begin{displaymath}
 H_2 = H_0 + H_d + \frac{1}{2} [{\cal S}, H_{\rm int}].
\end{displaymath}
The generator ${\cal S}$ satisfies the equation $[H_d + H_0, {\cal S}] =
H_{\rm int}$ and has the form \cite{muhlschlegel-1968-1,isaev-2013-1}:
\begin{displaymath}
 {\cal S} = \frac{v}{\sqrt{N}} \sum \frac{1}{E^\prime_d - E_d}
 \cd_{k a \sigma} P^\prime_d d_{a \sigma} P_d - {\rm h.c.}
\end{displaymath}
Here the summation is over all variable not present in the l.h.s. The
operators $P_d$ are projectors on the triangle states with energy $E_d$,
and we neglected the electron bandwidth in the energy denominators.

The matrix elements of the commutator in $H_2$ can be computed as
\begin{widetext}
 \begin{align}
  \langle a_1 \vert [{\cal S}, & H_{\rm int}] \vert a_2 \rangle =
  \frac{v^2}{N} \sum \frac{1}{E_d^\prime - E_d} \langle a_1 \vert
  (\cd_{k a \alpha} P_d^\prime d_{a \alpha} P_d) (\dd_{b \beta}
  c_{p b \beta}) - (\dd_{b \beta} c_{p b \beta}) (\cd_{k a \alpha} P_d^\prime
  d_{a \alpha} P_d) \vert a_2 \rangle + {\rm h.c.} = \nonumber \\
  & = \frac{v^2}{N} \sum \biggl[ \frac{1}{E_d(2, 0) - E_d(3, 1/2)}
  \cd_{k a \alpha}
  c_{p b \beta} \langle a_1 \vert d_{a \alpha} \dd_{b \beta} \vert a_2
  \rangle - \frac{1}{E_d(1, 1/2) - E(2, 0)} c_{p b \beta} \cd_{k a \alpha}
  \langle a_1 \vert \dd_{b \beta} d_{a \alpha} \vert a_2 \rangle \biggr] +
  {\rm h.c.}, \nonumber
 \end{align}
\end{widetext}
where $E_d(2, 0) - E_d(3, 1/2) = -\bigl( 2 U - \epsilon_d + \frac{3}{4} J_H
\bigr)$ and $E_d(1, 1/2) - E_d(2, 0) = \epsilon_d - U + \frac{3}{4} J_H$.
We will focus on a special case $\epsilon_d = \frac{3}{2} U$ when
$E_d(3, 1/2) - E_d(2, 0) = E_d(1, 1/2) - E_d(2, 0) = \Delta$.

Since the states $\vert a_{1, 2} \rangle$ in Eq. (2) are singlets, the matrix
elements $\langle a_1 \vert \cdots \vert a_2 \rangle$ satisfy the property
\begin{displaymath}
 \langle a_1 \vert d_{a \alpha} \dd_{b \beta} \vert a_2 \rangle =
 \delta_{\alpha \beta} \langle a_1 \vert d_{a \ua} \dd_{b \ua} \vert a_2
 \rangle = \delta_{\alpha \beta} \langle a_1 \vert d_{a \da} \dd_{b \da}
 \vert a_2 \rangle
\end{displaymath}
and similarly for $\langle a_1 \vert \dd_{b \beta} d_{a \alpha} \vert a_2
\rangle$. The spin-polarized elements are given by
\begin{align}
 \langle a_1 \vert d_{a \ua} \dd_{b \ua} \vert a_2 \rangle = &
 \delta_{a b} \delta_{a_1 a_2} \delta_{a_2 a} - \frac{1}{2}
 \delta_{a_1 a} \delta_{a_2 b} (1 - \delta_{a b}), \nonumber \\
 \langle a_1 \vert \dd_{b \ua} d_{a \da} \vert a_2 \rangle = &
 \frac{1}{2} \biggl[ \delta_{a b} \delta_{a_1 a_2} (1 \! - \! \delta_{a_2 a}) +
 (1 \! - \! \delta_{a b}) \delta_{a_1 a} \delta_{a_2 b} \biggr]. \nonumber
\end{align}
Using these expressions and neglecting the constant term that arises from
$c_{p b \beta} \cd_{k a \alpha} = \delta_{a b} \delta_{k p}
\delta_{\alpha \beta} - \cd_{k a \alpha} c_{p b \beta}$, we arrive at the
Kondo impurity model
\begin{align}
 H_{\rm ef} = & \sum_k \varepsilon_{k \sigma} \cd_{k a \sigma}
 c_{k a \sigma} - \frac{V_\perp}{N} \sideset{}{^\prime} \sum_{k p}
 \fd_{a^\prime} f_a \cd_{k a^\prime \sigma} c_{p a \sigma} - \nonumber \\
 & - \frac{V_\|}{N} \sum_{k p} \fd_a f_a \cd_{k a \sigma} c_{p a \sigma}
 \nonumber
\end{align}
with $V_\| = \frac{3 v^2}{2 \Delta}$ and $V_\perp = -\frac{v^2}{2 \Delta}$,
from which Eq. (3) is recovered by adding a triangle position $i$,
and replacing in the last two terms the summation over $k$ and $p$ with that
over $i$:
\begin{displaymath}
 \frac{1}{N} \sum_{k p} \cd_{k a^\prime \sigma} c_{p a \sigma} \to
 \sum_i \cd_{i a^\prime \sigma} c_{i a \sigma}.
\end{displaymath}
Note the term $H_d$ gives only a constant energy shift, since $\langle a_1
\vert H_d \vert a_2 \rangle = E_d(2, 0) \delta_{a_1 a_2}$, and therefore was
dropped.
The Hamiltonian $H_{\rm ef}$ can be reduced to a single-channel form using
the same arguments as in the main text. Hence, in the following we will
consider its spin-polarized version, omit the spin index $\sigma$ and replace
$\varepsilon_{k \sigma} \to \epsilon_k$:
\begin{equation}
 H_{\rm ef} \to \sum_k \epsilon_k \cd_{k a} c_{k a} - \frac{1}{N} \sum_{k p}
 V_{a b} \fd_a f_b \cd_{k a} c_{p b}
 \label{eq:kim-3qd}
\end{equation}
with $V_{a b} = V_\perp (1 - \delta_{a b}) + V_\| \delta_{a b}$.

Finally we note that an Anderson model similar to Eq. \eqref{eq:aim-3qd} was
studied in the context of tunneling through triple quantum dots
\cite{kuzmenko-2006-1,lopez-2013-1}, solid-state spin entanglers
\cite{sarago-2003-1}, and Kondo physics in frustrated geometries
\cite{ingersent-2005-1,mitchell-2013-1}, but to our knowledge has never been
generalized to a lattice.

\section{Physical properties of the $SU(3)$ KL}

In this section we elaborate on the discussion in the main text regarding
physical properties of the KL model (3). First, we will consider the limit of a
single $SU(3)$ impurity in order to identify parameter regimes relevant for
Kondo screening. Based on the obtained intuition, we then generalize the
hybridization mean-field approach \cite{coleman-2007-1} to handle the $SU(3)$
KL model in Eq. (3).

\subsection{Poor man scaling for a single impurity}

We will apply the perturbative (in $V_\|$ and $V_\perp$) renormalization group
(RG), i.e. Anderson's poor man scaling described in Ref.
\onlinecite{hewson-1997-1} to the Kondo impurity model \eqref{eq:kim-3qd}. For
simplicity, we assume that the itinerant band has a constant density of states
$\rho_0$ for $\epsilon \in [-D, D]$. Notice that system is not assumed to be
1D.

The perturbative RG approach is a series of consecutive Schrieffer-Wolff
transformations that remove high-energy states near the band edges
$[D - \delta D, D]$ and $[-D, -D + \delta D]$ with $\delta D > 0$, $\delta D
\ll D$. It is assumed that there are no electrons (holes) in the first (second)
interval. We will label states inside (outside) these intervals with momenta
$q$, $q^\prime$ ($k$, $p$), etc.

The impurity model \eqref{eq:kim-3qd} can be cast in the form
\begin{align}
 H_{\rm ef} = & \sum_k \epsilon_k \cd_{k a} c_{k a} + \sum_q \epsilon_q
 \cd_{q a} c_{q a} - \nonumber \\
 & - \frac{1}{N} \sum_{k p} V_{a b} \fd_b f_a \cd_{k b}
 c_{p a} - \frac{1}{N} \sum_{q^\prime q} V_{a b} \fd_b f_a \cd_{q^\prime b}
 c_{q a} - \nonumber \\
 & - \frac{1}{N} \sum_{p q} V_{a b} \fd_b f_a (\cd_{p b} c_{q a} + \cd_{q a}
 c_{p b}), \nonumber
\end{align}
where the first two lines contain diagonal terms and the last line is an
off-diagonal operator $H_1 = -\frac{1}{N} \sum_{p q} V_{a b} \fd_b f_a
(\cd_{p b} c_{q a} + \cd_{q a} c_{p b})$ that transfers electrons to and from
near the band edges. Note that $H_1 \sim \delta D$ because of the
$q$-summation, and we can neglect the term with $\sum_{q^\prime q} (\cdots)
\sim (\delta D)^2$. The ``unperturbed'' Hamiltonian for the band edges can be
taken simply as $H_0 = \sum_q \epsilon_q \cd_{q a} c_{q a}$ with $\epsilon_q
\approx \pm D$.

The transformation $\e^{\cal S} (H_0 + H_1) \e^{-{\cal S}} \approx H_0 +
\frac{1}{2} [{\cal S}, H_1]$ removes $H_1$, and the r.h.s. of this expression
provides a correction to the remaining Hamiltonian which now does not contain
the band-edge degrees of freedom. All coupling constants in this new
Hamiltonian are defined at an energy scale $D - \delta D$. The generator
${\cal S}$ is determined by the equation $H_1 = [H_0, {\cal S}]$ and can be
symbolically written as
\begin{displaymath}
 {\cal S} = -\frac{1}{N} \sum \frac{V_{a b}}{E^\prime_0 - E_0} \fd_b f_a
 \cd_{p b} P^\prime_0 c_{q a} P_0 - {\rm h.c.}
\end{displaymath}
Here $E_0$ ($E_0^\prime$) are eigenstates of $H_0$, and $P_0$ ($P_0^\prime$)
corresponding projectors. The summation extends over all relevant variables.

\begin{figure}[t]
 \begin{center}
  \includegraphics[width = \columnwidth]{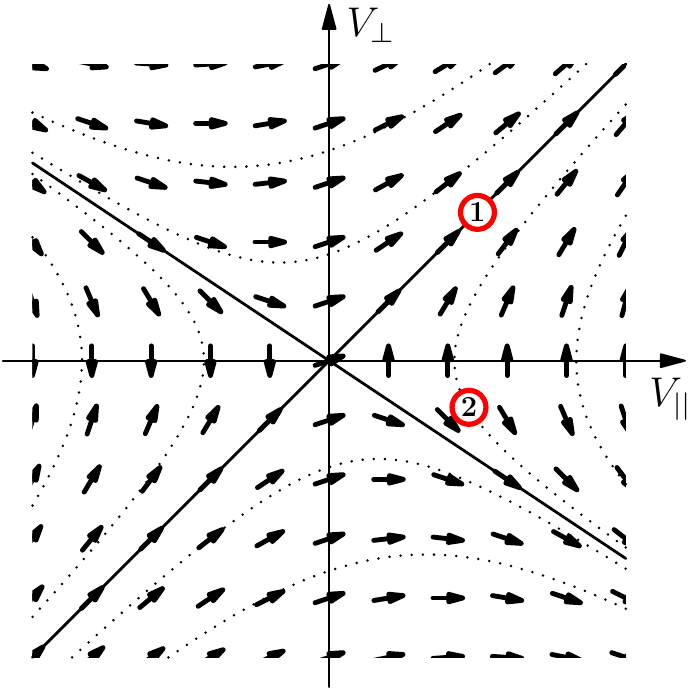}
 \end{center}
 \caption{
  Renormalization group flow defined by Eq. \eqref{eq:rg-flow}. Solid lines are
  the trajectories \eqref{eq:rg-darboux}. Labels 1 and 2 correspond to the two
  cases studied in the text, and represent the model \eqref{eq:kim-3qd} with
  $V_\perp = V_\|$ and $V_\perp = -\frac{2}{3} V_\|$ respectively.
 }
 \label{fig:sfig2}
\end{figure}

A straightforward calculation of the commutator $[{\cal S}, H_1]$
yields the RG equations:
\begin{align}
 V_\perp(D - \delta D) = & V_\perp(D) + V_\perp (V_\perp + 2 V_\|) \,\,
 \frac{\rho_0 \delta D}{D}, \nonumber \\
 V_\|(D - \delta D) = & V_\|(D) + 3 V_\perp^2 \,\, \frac{\rho_0 \delta D}{D},
 \nonumber
\end{align}
or in a differential form:
\begin{equation}
 \frac{d V_\perp}{d \ln D} = -\rho_0 V_\perp (V_\perp + V_\|); \,\,\,
 \frac{d V_\|}{d \ln D} = -3 \rho_0 V_\perp^2.
 \label{eq:rg-flow}
\end{equation}
The flow trajectories (integrals of motion) are obtained by excluding $\ln
D$ from these expressions, i.e. $d V_\| / d V_\perp = 3 V_\perp / (2 V_\| +
V_\perp)$. This is a Darboux equation \cite{zaitsev-2002-1}, whose solution is
\begin{equation}
 (V_\perp - V_{\vert \vert})^3 \biggl( V_\perp + \frac{2}{3} V_{\vert \vert}
 \biggr)^2 = {\rm const}.
 \label{eq:rg-darboux}
\end{equation}
It is easy to find the RG flow in two special regimes: (1) $V_\perp = V_\|$
[full $SU(3)$ symmetry] and (2) $V_\perp = -\frac{2}{3} V_\|$, both
corresponding to ${\rm const} = 0$ in Eq. \eqref{eq:rg-darboux}. Assuming that
$V_\| = V^{(0)}_\|$ for some fixed value of $D = D_0$, in the first case we
have: $V_\|(D) = V^{(0)}_\| / \bigl[ 1 - 3 \rho_0 V^{(0)}_\| \ln \frac{D_0}{D}
\bigr]$. In the second case the solution is $V_\|(D) = V^{(0)}_\| / \bigl[ 1 -
\frac{4}{3} \rho_0 V^{(0)}_\| \ln \frac{D_0}{D} \bigr]$. Upon reducing the
bandwidth $D < D_0$, $V_\|$ always increases. Hence, for $V^{(0)}_\| > 0$, the
flow is always towards strong coupling indicating a Kondo-screened GS. The
scaling trajectories for a general case are shown in Fig. \ref{fig:sfig2}.

\subsection{The hybridization mean-field approach}

The RG flow in Fig. \ref{fig:sfig2} suggests that for $V_\| > 0$, the impurity
model \eqref{eq:kim-3qd} always exhibits the Kondo screening. Similar to the
$SU(2)$ case \cite{coleman-2007-1}, one may expect that the same is true for
the lattice model (3). Here we demonstrate this by using a modified slave boson
(hybridization) mean-field approach that replaces pseudo-fermions $f_{i a}$ in
Eq. (3) with canonical fermions at unit filling:
\begin{equation}
  n^f_i = 1 \to \frac{1}{N} \sum_i \langle n^f_i \rangle = 1,
  \label{eq:mf_constraint}
\end{equation}
where $n^f_i = \fd_{i a} f_{i a}$ and $N$ is the number of lattice
sites.

The correct slave boson fields can be identified by diagonalizing the
interaction part in Eq. (3) on a single site assuming that there is one $f$-
and $c$-fermion. There are nine eigenstates
\begin{align}
 \vert \psi_0 \rangle = & \frac{1}{\sqrt{3}} \sum_a \fd_{i a} \cd_{i a} \vac;
 \label{eq:states} \\
 \vert \psi_{1, 2} \rangle = & \frac{1}{\sqrt{3}} (\fd_{i 1} \cd_{i 1} +
 \omega^{\pm 1} \fd_{i 2} \cd_{i 2} + \omega^{\mp 1} \fd_{i 3} \cd_{i 3}) \vac;
 \nonumber \\
 \vert \psi_{3 \ldots 8} \rangle = & \fd_{i a} \cd_{i b} \vac \quad
 {\rm with} \,\,\, a \neq b \nonumber
\end{align}
with energies $E_0 = -(V_\| + 2 V_\perp)$, $E_1 = E_2 = -(V_\| - V_\perp)$,
$E_{3 \ldots 8} = 0$, and $\omega = \e^{2 \pi \ii / 3}$.

The states $\vert \psi_{1, 2} \rangle$ transform according to the
two-dimensional representation of the group $C_{3 v}$ \cite{bir-1974-1}. The
symmetry operations from this group are shown in the inset of Fig.
\ref{fig:sfig1}, and their matrices in the basis $\vert \psi_{1, 2} \rangle$
are given by:
\begin{displaymath}
 C_3 \! = \!
 \begin{pmatrix}
  \omega & 0 \\
  0 & \omega^*
 \end{pmatrix}
 \! , \, \,
 \sigma^1_v \! = \!
 \begin{pmatrix}
  0 & 1 \\
  1 & 0
 \end{pmatrix}
 \! , \, \,
 \sigma^2_v \! = \!
 \begin{pmatrix}
  0 & \omega \\
  \omega^* & 0
 \end{pmatrix}
 \! , \, \,
 \sigma^3_v = \bigl( \sigma^2_v \bigr)^*
\end{displaymath}
The state $\vert \psi_0 \rangle$ is invariant under all operations from
$C_{3 v}$.

Each of the states \eqref{eq:states} can be associated with a Schwinger boson
${\hat \chi_{i n}}$ (for a fixed $i$), and we can formally write the Kondo term
as
\begin{align}
 H_K = & \sum_i \Lambda^{a^\prime b^\prime}_{a b} \fd_{i a^\prime}
 \cd_{i b^\prime} c_{i b} f_{i a} = \nonumber \\
 & = \sum_{i n} E_n \langle a^\prime b^\prime \vert n \rangle \langle n \vert a
 b \rangle \fd_{i a^\prime} \cd_{i b^\prime} c_{i b} f_{i a} = \nonumber \\
 & = \sum_{i n} E_n {\hat \chi}^\dag_{i n} {\hat \chi}_{i n}. \nonumber
\end{align}
In our case the slave bosons are actually pairing fields:
\begin{align}
 {\hat \chi}_{i 0} = & \frac{1}{\sqrt{3}} \sum_a f_{i a} c_{i a};
 \label{eq:slave-bosons} \\
 {\hat \chi}_{i 1, 2} = & \frac{1}{\sqrt{3}} (f_{i 1} c_{i 1} + \omega^{\mp 1}
 f_{i 2} c_{i 2} + \omega^{\pm 1} f_{i 3} c_{i 3}) \nonumber
\end{align}
Other $\chi_{i n}$ with $n = 3 \ldots 8$ do not appear due to vanishing
eigenvalues $E_{3 \ldots 8}$.

We assume that the system is homogeneous so all averages are $i$-independent,
and perform a Hartree-Fock-Bogoliubov decoupling in two ``channels'': (i)
Hybridization channel, by condensing the $\chi$-bosons, i.e. considering the
averages $\chi_n = \langle {\hat \chi_{i n}} \rangle$, and (ii) magnetization
channel, defined by the traceless parts of $\langle \fd_{i a} f_{i b} \rangle$
and $\langle \cd_{i a} c_{i b} \rangle$. The latter can be expanded in terms of
Gell-Mann matrices $\lambda^l$ with $l = 1 \ldots 8$:
\begin{align}
 \langle \fd_{i a} f_{i b} \rangle = & \sum_l \lambda^l_{a b}
 m^f_l + \frac{1}{3} \delta_{a b}, \nonumber \\
 \langle \cd_{i a} c_{i b} \rangle = & \sum_l \lambda^l_{a b}
 m^c_l + \frac{n^c}{3} \delta_{a b}. \nonumber
\end{align}
Our convention for the Gell-Mann matrices is:
\begin{align}
 \lambda^1 = &
 \begin{pmatrix}
  0 & 1 & 0 \\
  1 & 0 & 0 \\
  0 & 0 & 0
 \end{pmatrix}
 \! , \,\,\,
 \lambda^2 =
 \begin{pmatrix}
  0 & -\ii & 0 \\
  \ii & 0 & 0 \\
  0 & 0 & 0
 \end{pmatrix}
 \! , \,\,\,
 \lambda^3 =
 \begin{pmatrix}
  1 & 0 & 0 \\
  0 & -1 & 0 \\
  0 & 0 & 0
 \end{pmatrix}
 \! , \nonumber \\
 \lambda^4 = &
 \begin{pmatrix}
  0 & 0 & 1 \\
  0 & 0 & 0 \\
  1 & 0 & 0
 \end{pmatrix}
 \! , \,\,\,
 \lambda^5 =
 \begin{pmatrix}
  0 & 0 & -\ii \\
  0 & 0 & 0 \\
  \ii & 0 & 0
 \end{pmatrix}
 \! , \,\,\,
 \lambda^6 =
 \begin{pmatrix}
  0 & 0 & 0 \\
  0 & 0 & 1 \\
  0 & 1 & 0
 \end{pmatrix}
 \! , \nonumber \\
 \lambda^7 = &
 \begin{pmatrix}
  0 & 0 & 0 \\
  0 & 0 & -\ii \\
  0 & \ii & 0
 \end{pmatrix}
 \! , \,\,\,
 \lambda^8 = \frac{1}{\sqrt{3}}
 \begin{pmatrix}
  1 & 0 & 0 \\
  0 & 1 & 0 \\
  0 & 0 & -2
 \end{pmatrix}
 \! . \nonumber
\end{align}
The mean-field Hamiltonian has the form:
\begin{align}
 H_{\rm MF} = & \sum_k \bigl[ (\epsilon_k - \mu_c) \delta_{a b} -
 M^f_{a b} \bigr] \cd_{k a} c_{k b} + \nonumber \\
 & + \sum_k \bigl[ \mu_f \delta_{a b} - M^c_{a b} \bigr] \fd_{-k a} f_{-k b} -
 \nonumber \\
 & - \sum_k \bigl[ V_a \cd_{k a} \fd_{-k a} + V^*_a f_{-k a} c_{k a} \bigr] =
 \nonumber \\
 & = \sum_k
 \begin{pmatrix}
  \cd_{k a} & f_{-k a}
 \end{pmatrix}
 {\cal H}_k
 \begin{pmatrix}
  c_{k b} \\
  \fd_{-k b}
 \end{pmatrix}
 \nonumber
\end{align}
with
\begin{displaymath}
 {\cal H}_k = 
 \begin{pmatrix}
  (\epsilon_k - \mu_c) \delta_{a b}- M^f_{a b} & -V_a \delta_{a b} \\
  -V^*_a \delta_{a b} & -\mu_f \delta_{a b} + M^c_{b a}
 \end{pmatrix}
 \! .
\end{displaymath}
Here summations over color indices are implicit and the $k$-summation extends
over the entire Brillouin zone. The chemical potential $\mu_f$ enforces the
constraint \eqref{eq:mf_constraint}, while $\mu_c$ fixes an average number of
the itinerant fermions:
\begin{displaymath}
 \frac{1}{N} \sum_{k a} \langle \cd_{k a} c_{k a} \rangle = n^c.
\end{displaymath}
Finally, the mean fields are $M^{c, f}_{a b} = \bigl[ V_\| \delta_{a b} +
V_\perp (1 - \delta_{a b}) \bigr] \sum_l \lambda^l_{a b} m^{c, f}_l$, and
$\sqrt{3} V_a = (V_\| + 2 V_\perp) \chi_0 + (V_\| - V_\perp) [\chi_1 \omega +
\chi_2 \omega^*]$. The Hamiltonian $H_{\rm MF}$ can be diagonalized by a
Bogoliubov transformation \cite{blaizot-1986-1}
\begin{displaymath}
 \begin{pmatrix}
  c_{k a} \\
  \fd_{-k a}
 \end{pmatrix}
 = \sum_s^{n_+(k)} X^s(k) \, \gamma_{k s} + \sum_t^{n_-(k)} Y^t(k) \,
 \beta^\dag_{k t}
\end{displaymath}
with $n_+$ [$n_-$] is the number of positive [negative] eigenvalues $E_s(k)$
[$-E_t(k)$] ($E_{s, t} > 0$) of ${\cal H}_k$. These numbers of course depend
on $k$ and the mean-field parameters. The fermionic quasiparticles
$\gamma_{k s}$ and $\beta_{k t}$ define the particle-hole excitations and have
only positive energies:
\begin{displaymath}
 H_{\rm MF} \to \sum_k \biggl[ \sum_s^{n_+(k)} E_s(k) \gamma^\dag_{k s}
 \gamma_{k s} + \sum_t^{n_-(k)} E_t(k) \beta^\dag_{k t} \beta_{k t} \biggr].
\end{displaymath}
We also note a useful relation
\begin{align}
 n^c - n^f = & \,\, n^c - 1 = \nonumber \\
 = \frac{1}{N} & \sum_k \biggl[ \sum_s^{n_+(k)} n^\gamma_{k s} -
 \sum_t^{n_-(k)} n^\beta_{k t} + n_-(k) - 3 \biggr]. \nonumber
\end{align}
Once the quasiparticles are known, we can self-consistently determine the order
parameters $\chi_n$ and $m^{c, f}_l$.

\subsection{Stability of Kondo-screened phases against lead-mixing
perturbations $H_{\rm mix}$}

In the main text we only briefly mentioned effects of the lead-mixing term
$H_{\rm mix}$ in Eq. (1) on the heavy-fermion phases in Figs. 2 and 3.
Here we provide a more detailed discussion of this issue.

The simplest form of $H_{\rm mix}$ compatible with $C_{3 v}$ symmetry of the
TTL corresponds to hopping of itinerant and local fermions around a triangle:
\begin{displaymath}
 H_{\rm mix} = -t^c_\Delta \sum_{\langle a b \rangle, k \sigma}
 \cd_{k a \sigma} c_{k b \sigma} - t^d_\Delta
 \sum_{\langle a b \rangle, i \sigma} \dd_{i a \sigma} d_{i b \sigma} +
 {\rm h.c.},
\end{displaymath}
where $t^{c, d}_\Delta$ are hopping amplitudes for itinerant and local
fermions, and $\langle a b \rangle$ is a triangle edge (nearest-neighbor link).
This Hamiltonian is diagonalized by introducing a transverse momentum $\lambda
= 0, \pm \frac{2 \pi}{3}$:
\begin{displaymath}
 H_{\rm mix} = \sum_{k \lambda \sigma} \xi^c_\lambda \cd_{k \lambda \sigma}
 c_{k \lambda \sigma} + \sum_{i \lambda \sigma} \xi^d_\lambda
 \dd_{i \lambda \sigma} d_{i \lambda \sigma}.
\end{displaymath}
with $\xi^{c, d}_\lambda = -2 t^{c, d}_\Delta \cos \lambda$,
$c_{k \lambda \sigma} = \frac{1}{\sqrt{3}} \sum_a \e^{-\ii \lambda a}
c_{k a \sigma}$ and similarly for $d_{i \lambda \sigma}$.
In all other terms in $H^{TTL}$ (1) we can simply replace the lead index $a$
with $\lambda$ and repeat calculations leading to the KL model (3) using the
dressed fermions $c_{k \lambda \sigma}$ and $d_{i \lambda \sigma}$.
Note, that the replacement $a \to \lambda$ also needs to be performed in the
expression (2) for the VB singlets.
The $d$-term in $H_{\rm mix}$ is diagonal in the basis of the VB states. For
example, $\sum_\sigma \dd_{i \lambda \sigma} d_{i \lambda \sigma} \vert 1
\rangle_i = (\xi^d_2 + \xi^d_3) \vert 1 \rangle_i = - \xi^d_1 \vert 1
\rangle_i$ because $\sum_\lambda \xi^{c, d}_\lambda = 0$, so in general
$\sum_\sigma \dd_{i \lambda \sigma} d_{i \lambda \sigma} \vert a \rangle_i =
-\xi^d_a \vert a \rangle_i$.
Hence, the net effect of $H_{\rm mix}$ amounts to introducing Zeeman splittings
$h^c_\lambda = \xi^c_\lambda$ for $c$-fermions and $h^d_\lambda =
-\xi^d_\lambda$ for the local $SU (3)$ spins:
\begin{displaymath}
 \delta H^{TTL}_{\rm ef} = \sum_{i \lambda} (h^c_\lambda \cd_{i \lambda}
 c_{i \lambda} + h^d_\lambda \fd_{i \lambda} f_{i \lambda}).
\end{displaymath}

In the context of the $SU (2)$ spin Kondo physics it is known
\cite{hewson-1997-1} that a sufficiently large magnetic field (of the order of
the Kondo temperature) suppresses the Kondo effect.
The same will happen in our case: The lead-mixing hopping amplitudes
$t^{c, d}_\Delta$ play the role of an applied field, and when they become large
enough the local GS degeneracy on each triangular plaquette will be lifted
which in turn will destroy the frustrated heavy-fermion phases in Figs. 2 and
3.

\bibliographystyle{apsrev4-1}
\bibliography{references}